\RequirePackage{fix-cm}
\documentclass[twocolumn,epjc3]{svjour3}  

\smartqed  
\RequirePackage{graphicx}
\usepackage{caption}
\usepackage{amsmath}
\usepackage{subfig}
\usepackage{braket}
\usepackage{lipsum}
\begin{document}

\title{Measurements of neutron total and capture cross sections of $^{139}$La and evaluation of resonance parameters
}


\author{Shunsuke Endo\thanksref{e1,addr1,addr2}
        \and
        Shiori Kawamura\thanksref{addr2} 
         \and
        Takuya Okudaira\thanksref{addr2}
         \and
        Hiromoto Yoshikawa\thanksref{addr3}
         \and
        Gerard Rovira\thanksref{addr1}
         \and
        Atsushi Kimura\thanksref{addr1}
         \and
        Shoji Nakamura\thanksref{addr1}
         \and
        Osamu Iwamoto\thanksref{addr1}
         \and
        Nobuyuki Iwamoto\thanksref{addr1}
}

\thankstext{e1}{e-mail: endo.shunsuke@jaea.go.jp}


\institute{Japan Atomic Energy Agency, 2-4 Shirakata, Tokai 319-1195, Japan \label{addr1}
           \and
            Nagoya University, Furocho, Chikusa, Nagoya 464-8062, Japan \label{addr2}
           \and
          Osaka University, Ibaraki, Osaka 567-0047, Japan  \label{addr3}
}

\date{Received: date / Accepted: date}

\maketitle

\begin{abstract}
Neutron total and capture cross sections of Lanthanum(La)-139 were measured at the Accurate Ne-utron-Nucleus Reaction measurement Instrument (ANNRI) of the Materials and Life Science Experimental Facility (MLF) in the Japan Proton Accelerator Research Complex (J-PARC). The total cross section was largely different from that in evaluated libraries, such as JENDL-5, in the energy range from 80 to 900~eV. Resonance parameters for four resonances including one negative resonance were obtained using a resonance analysis code, REFIT. The resonance analysis revealed discrepancies in several resonance parameters with the evaluated libraries. Furthermore, the information about the scattering radius was also extracted from the results of the total cross section. The obtained scattering radius was larger than that recorded in the evaluated libraries.
\end{abstract}

\section{Introduction}
The cross sections and resonance parameters of lanthanum (La)-139 are significant for applications in a nuclear technology and studies in a fundamental nuclear physics. In applications in the nuclear technology, an activation detector on the basis of neutron capture reactions of $^{139}$La is applied to neutron dosimetry in the epi-thermal neutron energy~\cite{INDC}. Furthermore, $^{139}$La is produced as one of fission products (5\% fission yield~\cite{England}) in a nuclear reactor; it is often  used to measure the operating power distribution after the nuclear reactor shutdown because of the short half-life (1.678~d) of $^{140}$La produced by neutron capture of $^{139}$La~\cite{Priyada}. Therefore, the accurate cross sections are required.

In fundamental nuclear physics studies, La has been used to clarify an enhancement mechanism of parity violation in compound nuclear reactions. The 0.75-eV resonance of $^{139}$La gave the magnitude of the parity violation as large as 10\%~\cite{Bowman,Yuan}, which is a very large enhancement among various nuclei~\cite{Michel}. An "s-p mixing model" was proposed to explain the enhancement mechanism of the parity violation, which took into account of the mixing between s- and p-wave amplitudes in the entrance channel of the compound nuclear state~\cite{Flambaum}. Angular correlation terms of capture reactions which are important to verify the s-p mixing model have recently been measured for several nuclei, such as $^{139}$La, $^{117}$Sn and $^{131}$Xe~\cite{Okudaira,Okudaira2021,Yamamoto,Koga,Endo2022,Okudaira2023}. Resonance parameters are needed to theoretically interpret the results of these correlation terms based on the s-p mixing model because the cross section of the correlation terms depends on the resonance parameters. Since $^{139}$La is one of the most significant nuclei in the study of the s-p mixing model due to observing the large enhancement of the parity violation, the accurate resonance parameters are required.
 
The neutron cross sections and the resonance parameters of $^{139}$La have been reported from several measurements. Terlizzi et al. measured the capture cross section at the n\_TOF in CERN~\cite{nTOF}. They measured capture $\gamma$-rays with C$_6$D$_6$ detectors located at the 187.5-m flight length. The resonance parameters from 0.75-eV to 8970-eV resonances were determined using a resonance analysis code, SAMMY~\cite{SAMMY}. Shwe et al. measured the total cross section in the Argonne fast chopper at the CP-5 reactor~\cite{Shwe}. Boron-loaded liquid scintillators were placed at the flight length of 60 and 120 m. The resonance parameters from 0.75 eV to 10352 eV were derived by an area analysis. Harvey et al.~\cite{Harvey} and Alfimenkov et al.~\cite{Alfimenkov} also measured transmission and obtained the resonance parameters of the 0.75-eV resonance. The discrepancies between these past studies are found in the neutron widths of the 0.75-eV resonance (see Table~\ref{respara}). Furthermore, in the s-p mixing model, the resonance parameters are necessary for not only the p-wave but also s-waves which are mixed into the p-wave resonance. In the $^{139}$La case, the 0.75-eV p-wave resonance is considered to most strongly mix with negative resonance~\cite{Okudaira}, but the negative resonance parameters were not reported in those past studies. To properly restrict the negative resonance parameters, it is significant to measure both the neutron total and capture cross sections and to determine the resonance parameters by simultaneous fits to both cross sections. Consequently, in this paper, the neutron transmission and the capture yield were measured at the Accurate Neutron-Nucleus Reaction measurement Instrument (ANNRI) equipped at Beam Line 04 of the Materials and Life Science Experimental Facility (MLF) in the Japan Proton Accelerator Research Complex (J-PARC). The neutron total and capture cross sections were derived from the neutron transmission and capture yield, respectively, and the resonance parameters were evaluated using a resonance analysis code, REFIT~\cite{REFIT}.

\section{Experiment}
\begin{figure*}[htbp]
\centering
\includegraphics[clip,width=16cm]{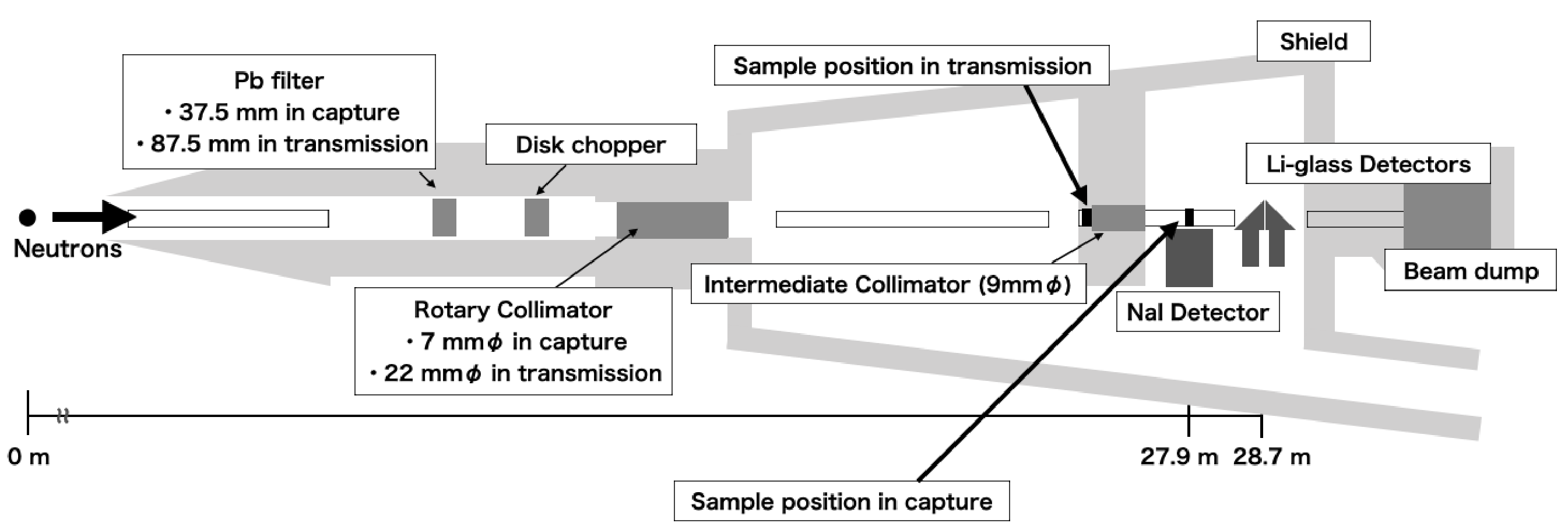}
\caption{\label{setup}Overview of ANNRI and experimental conditions for the collimators and filter.}
\end{figure*}

In MLF, pulsed protons are provided with a specific repetition rate, 25 Hz, and neutrons are produced by the proton-induced spallation reactions in the mercury target. The proton beam power was 800~kW, and the beam had a double-bunch structure~\cite{Endo} at the time of the present experiment. The time-of-flight (TOF) method was applied to determine the neutron energy. Figure~\ref{setup} shows an overview of ANNRI and experimental conditions for the collimators and filter. Neutron transmission was measured by Li-glass detectors installed at 28.7-m flight length. Two types of Li-glass detectors, $^{6}$Li enriched and $^{7}$Li enriched, were employed. The use of $^{7}$Li-glass detector was to subtract $\gamma$-ray backgrounds. The backgrounds caused by scattered neutrons and captured $\gamma$-rays in a sample were negligibly removed with the intermediate collimator placed between the sample and detectors. In capture cross section measurement, an NaI detector installed at 27.9-m flight length was used to detect $\gamma$-rays emitted via neutron capture reactions.

In the transmission measurement, a metallic natural La sample was used; its size was $40\ \times\ 40\ \times\ 10$ mm and areal density was $(2.698\pm0.013)\ \times\ 10^{-2}$~atoms/b. The sample contained a slightly amount of gadolinium as an impurity. The sample-out measurement, hereinafter called "blank", was also conducted to obtain the transmission. A gold (Au) sample was used to determine the flight length and initial time delay using the resonances of $^{197}$Au. In addition, the measurement inserting the black resonance filters (silver, manganese, cobalt, and indium) was performed to correct the difference of the detection efficiencies between $^{6}$Li and $^{7}$Li enriched detectors. A digitizer, CAEN V1720, was adopted to acquire the TOF and the pulse height with a list mode.

In the capture cross section measurement, another metallic natural La sample was used; its size was $\phi$10 $\times$ 0.175~mm and areal density was $(4.669\ \pm\ 0.093)\ \times\ 10^{-4}$~atoms/barn. The blank and carbon sample were measured to correct sample-independent and sample-scattered backgrounds. Samples of Au and boron were used to normalize the cross section and to obtain the incident neutron flux, respectively. The TOF and the pulse height were acquired by the digitizer CAEN V1724 with the list mode. Details of the cross section measurements at ANNRI are described in Ref.~\cite{Endo,Igashira,Gerard_Np,Kimura_Am,Terada_Am,Endo_Ta}.

\section{Analysis and Results}
\subsection{Total cross section}
\subsubsection{Derivation of transmission}
\begin{figure}[htbp]
\centering
\includegraphics[clip,width=8cm]{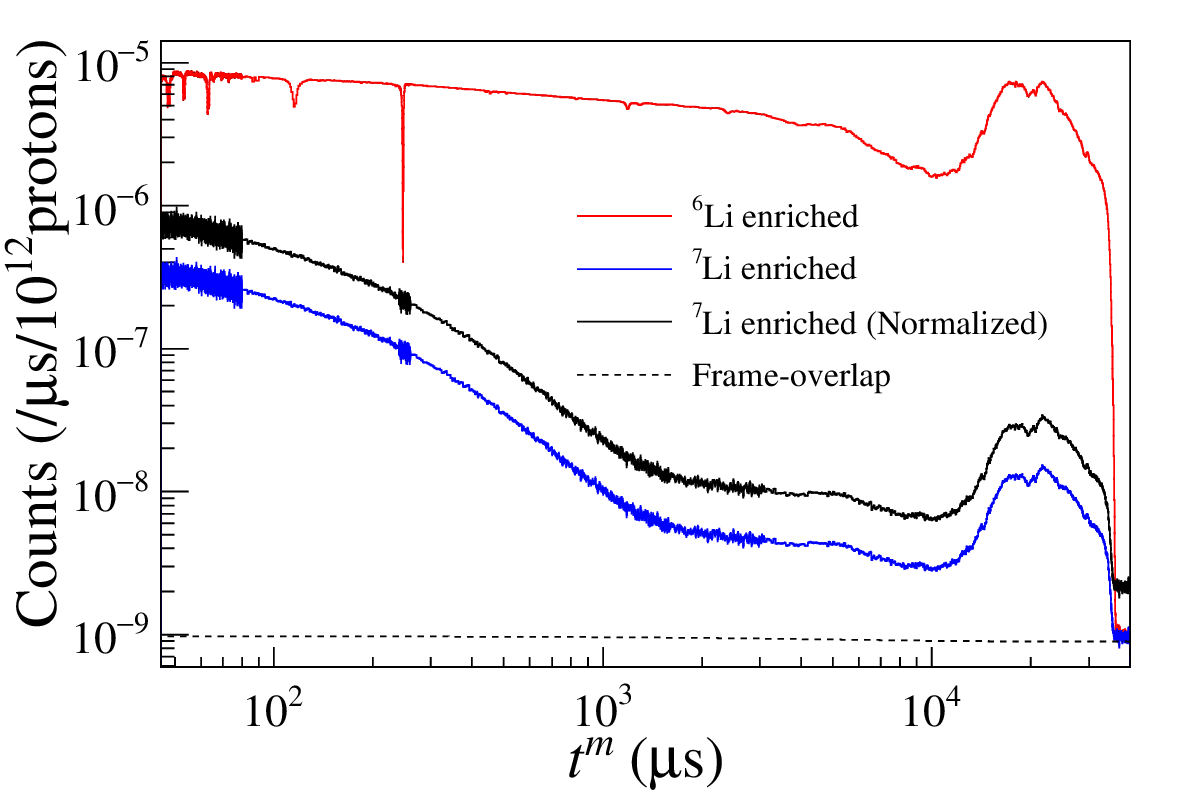}
\caption{\label{frameoverlapLi}TOF spectra as a function of TOF, $t^{m}$, obtained by $^{6}$Li and $^{7}$Li enriched detectors for La sample measurement. The normalized $^7$Li enriched spectrum means the $^7$Li enriched spectrum corrected by the difference of the detection efficiency. Frame-overlap backgrounds were estimated by fitting the spectra from 37 ms to 40 ms by an exponential function and constant.}
\end{figure}

A TOF spectrum was obtained by cutting the pulse height of $^{6}$Li(n,$\alpha$) reactions. The dead-time correction was performed by the extended dead-time model~\cite{Endo,Kimura_Am}, and the frame-overlap background, including a constant background, was estimated using the TOF region interrupted by the disk-chopper. Figure~\ref{frameoverlapLi} displays TOF spectra as a function of TOF, $t^{m}$, after dead-time corrections. The neutrons were blocked by disk-chopper after 37~ms. Thus, the region from 37~ms to 40~ms was fitted by an exponential function ($^{6}$Li enriched) and a constant ($^{7}$Li enriched), and the frame-overlap backgrounds were estimated by the fitting results extrapolating to the region after 40~ms. 

\begin{figure}[htbp]
\centering
\includegraphics[clip,width=8cm]{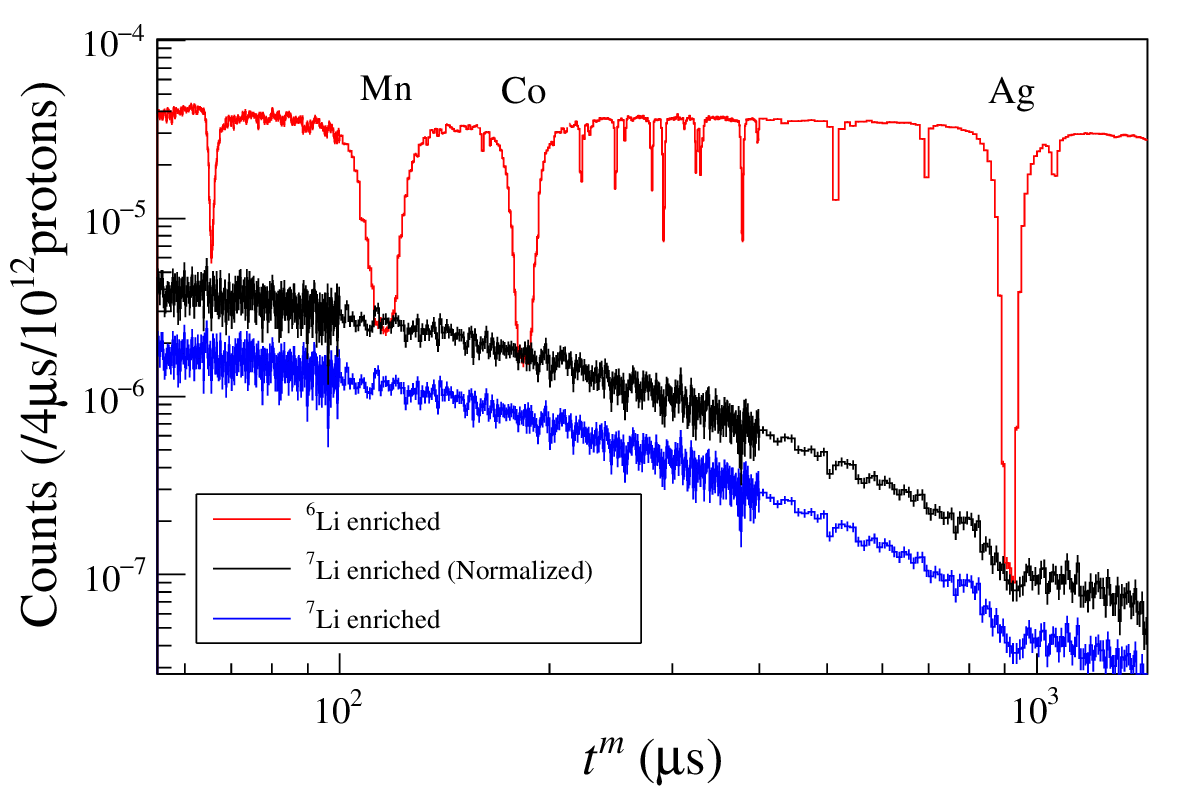}
\caption{\label{GS30norm}TOF spectra in the black resonance measurement. Difference of the detection efficiency between $^{6}$Li and $^{7}$Li enriched detectors was corrected to accord the spectrum at the black resonances.}
\end{figure}

The $\gamma$-ray background was removed by subtracting the TOF spectrum of the $^{7}$Li enriched detector from that of the $^{6}$Li enriched one. The TOF spectra in the black resonance measurement were used to correct the difference of the detection efficiency between these detectors and shown in Fig.~\ref{GS30norm}. The spectrum of the $^{7}$Li enriched detector was multiplied by $2.24\pm0.05$ to accord it with the bottom of the spectrum at the black resonances. The corrected $^7$Li enriched spectrum for the La measurement is simultaneously shown in Fig.~\ref{frameoverlapLi} as ``$^7$Li enriched (Normalized)".

\subsubsection{Results of total cross section}
\begin{figure}[htbp]
\centering
\includegraphics[clip,width=8.5cm]{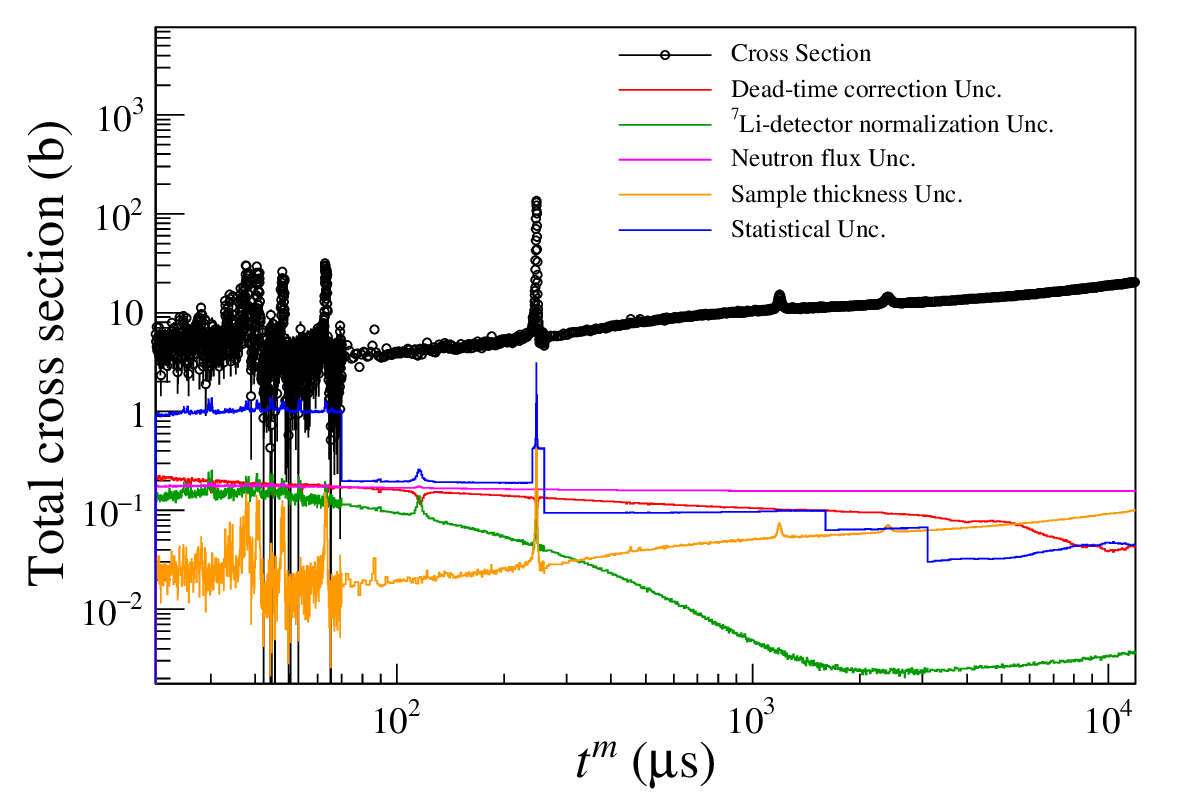}
\caption{\label{totcross1}Reduced total cross section with the statistical and various systematic uncertainties.}
\end{figure}

The transmission, $T$, was derived from the TOF spectrum of sample-in measurement divided by that of blank measurement. The total cross section was obtained as:
\begin{equation}
\tilde{\sigma}_\textrm{tot}(t^{m})=-\frac{1}{n_\textrm{La}}\ln T(t^{m}),
\end{equation}
were $n_\textrm{La}$ is the areal density. Here, this cross section includes the broadening effect by the resolution function and Doppler. The broadened cross section is hereafter referred to the reduced total cross section and the reduced capture cross section. The reduced total cross section is shown in Fig.~\ref{totcross1} with the statistical and various systematic uncertainties. The neutron flux uncertainty comes from the fluctuation of the intensity of the neutron beam due to the scattering by the air on the neutron beam line, which was estimated from the humidity and the atmospheric pressure~\cite{Endo_Ta}.

\begin{figure}[htbp]
\centering
\includegraphics[clip,width=8.5cm]{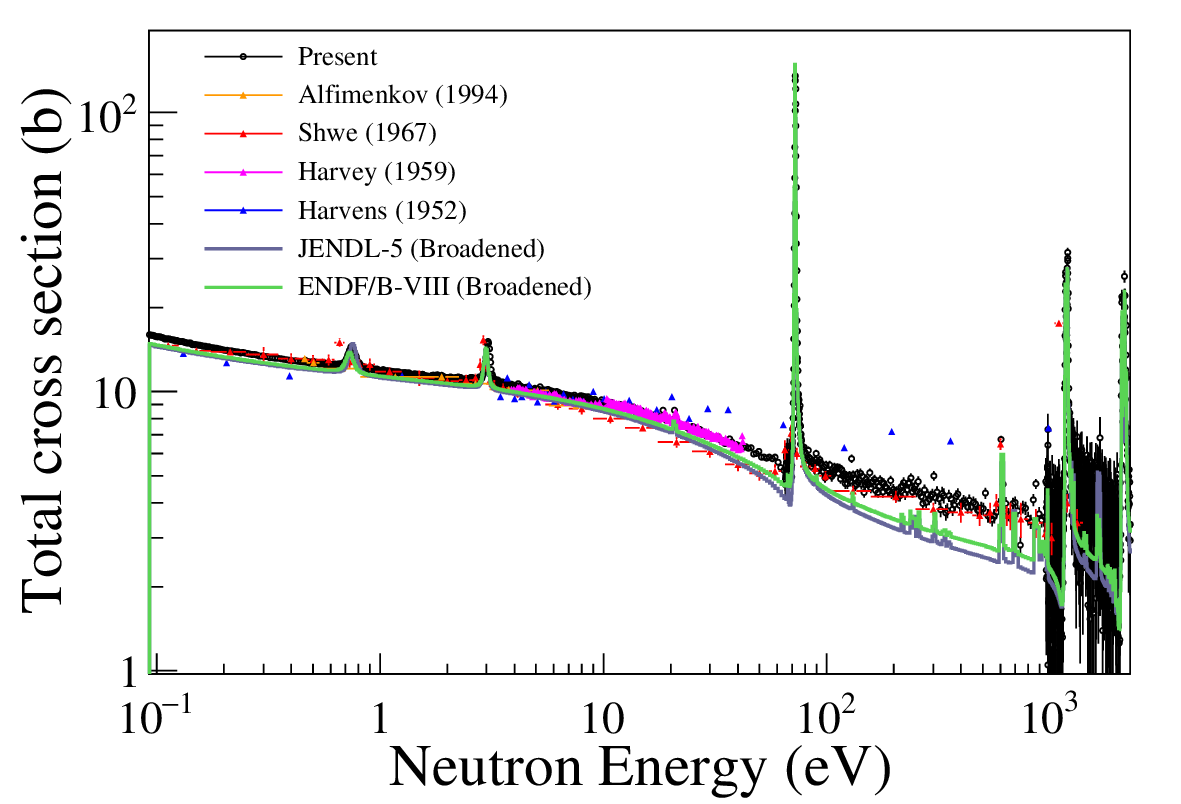}
\caption{\label{totcross2}Reduced total cross section with the past measurements and evaluated libraries. The cross sections of the evaluated libraries are broadened using the resolution function. The past measured data were taken from EXFOR~\cite{EXFOR}.}
\end{figure}

Figure~\ref{totcross2} represents the reduced total cross section comparing to the past measurements (taken from EXFOR~\cite{EXFOR}) and evaluated libraries broadened by the resolution function of ANNRI and Doppler effect. Since the resolution functions were different in each measurement, the shape of resonance has strong impacts. However, the region without resonance structure can be compared. The results are in good agreement with Harvey~\cite{Harvey} around 10~eV and with Shwe~\cite{Shwe} around a few 100~eV. On the other hand, the large discrepancy with the evaluated libraries, JENDL-5~\cite{JENDL} and ENDF/B-VIII.0~\cite{ENDF}, in the energy range from 80 to 900~eV, was found. 

\subsection{Capture cross section}
\subsubsection{Derivation of capture cross section}
The pulse height weighting technique (PHWT)~\cite{PHWT} was applied to the capture cross section analysis. The capture yield as a function of TOF, $t^{m}$, was calculated as:
\begin{equation}
    Y(t^{m})=\frac{\displaystyle \sum_{E_\gamma} W(E_\gamma)S(E_\gamma,t^{m})}{B_\textrm{n}+E_\textrm{n}},
\end{equation}
where $E_\gamma$ and $E_\textrm{n}$ are the $\gamma$-ray and neutron energies; $B_\textrm{n}$ is the neutron binding energy; $S(E_\gamma,t^{m})$ is the number of detected event of capture $\gamma$-rays; $W(E_\gamma)$ is the weighting function, and it was calculated using PHITS \cite{PHITS}. The obtained weighting function is as follows:
\begin{eqnarray}
  W(E_\gamma)&=&N\left\{a_0\left(\frac{E_\gamma}{25}\right)^{0.5}+a_1\left(\frac{E_\gamma}{25}\right)+a_2\left(\frac{E_\gamma}{25}\right)^{1.5}\right. \nonumber \\ &&\left.+a_2\left(\frac{E_\gamma}{25}\right)^{2}+a_3\left(\frac{E_\gamma}{25}\right)^{2.5}\right\}, \\
  \textrm{for La} \nonumber \\
  a_0&=&4.6621, \nonumber\\
  a_1&=&9.9137\times 10^{-2}, \nonumber \\
  a_2&=&1.8380\times 10^{-1}, \nonumber\\
  a_3&=&-1.1684\times 10^{-2}, \nonumber\\
  a_4&=&2.4360\times 10^{-4}, \nonumber \\
  \textrm{for Au} \nonumber \\
  a_0&=&9.1336, \nonumber \\
  a_1&=&-1.1183,\nonumber\\
  a_2&=&2.9693\times 10^{-1}, \nonumber\\
  a_3&=&-1.6273\times 10^{-2}, \nonumber\\
  a_4&=&3.1498\times 10^{-4}, \nonumber
\end{eqnarray}
and $N$ is the normalization factor. This factor was determined with the following constraint that the thermal capture yield of $^{197}$Au reproduces the thermal capture cross section in JENDL-5~\cite{JENDL}, 98.649~b. 

\begin{figure}[htbp]
\centering
\includegraphics[clip,width=8.5cm]{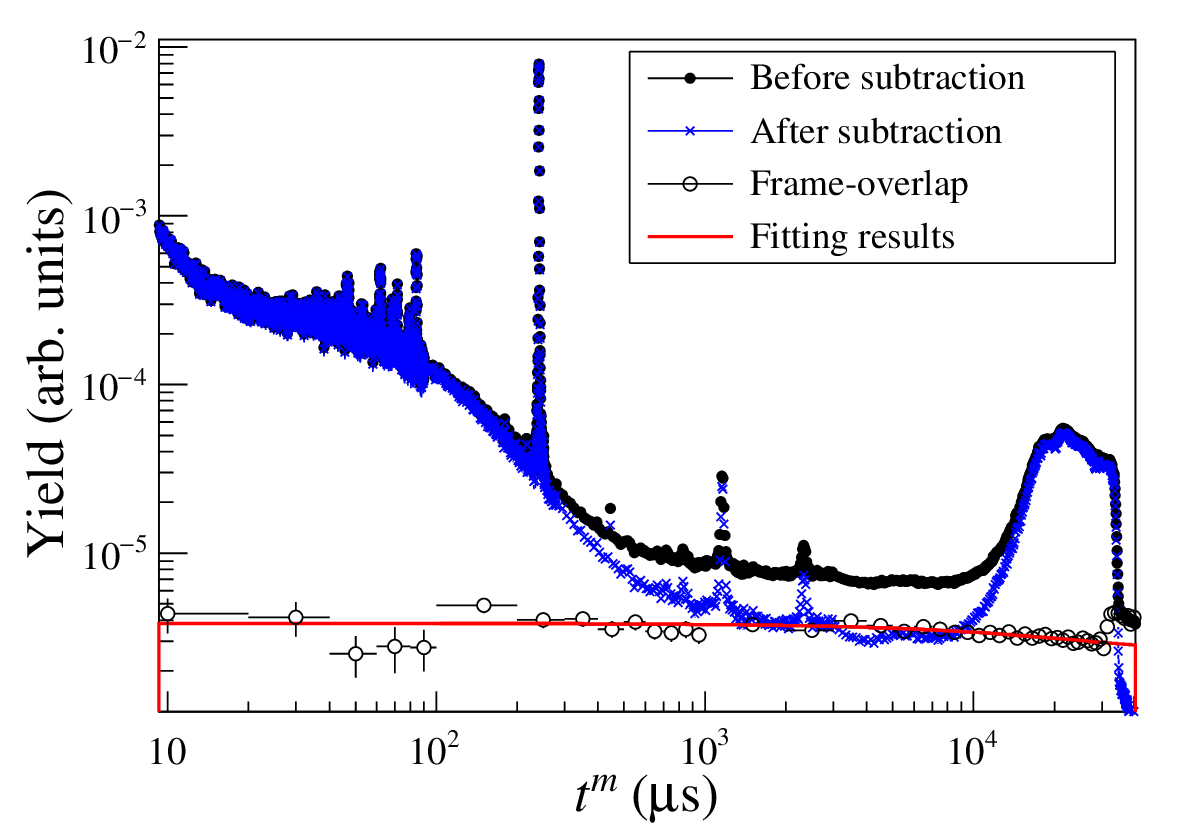}
\caption{\label{frameoverlapNaI}Yields as a function of $t^\textrm{m}$ before and after frame-overlap subtraction with the frame-overlap yield and fitting results.}
\end{figure}

The extended dead-time model was applied to the dead-time correction~\cite{Kimura_Am,Endo}. The frame-overlap background was obtained from the $\gamma$-ray counts when protons were not provided to MLF. In J-PARC, the protons are provided to two facilities, MLF and Main Ring (MR), with the specific allocation rate, MLF : MR =126 : 4. Therefore, when protons are provided to MR, a yield after 40 ms reflects the frame-overlap background. Figure~\ref{frameoverlapNaI} displays the yield multiplied by 126 together with a yield when protons were provided to the MLF. The frame-overlap background was subtracted from the capture yield using the fitting results performed with an exponential function.

\begin{figure}[htbp]
\centering
\includegraphics[clip,width=8.5cm]{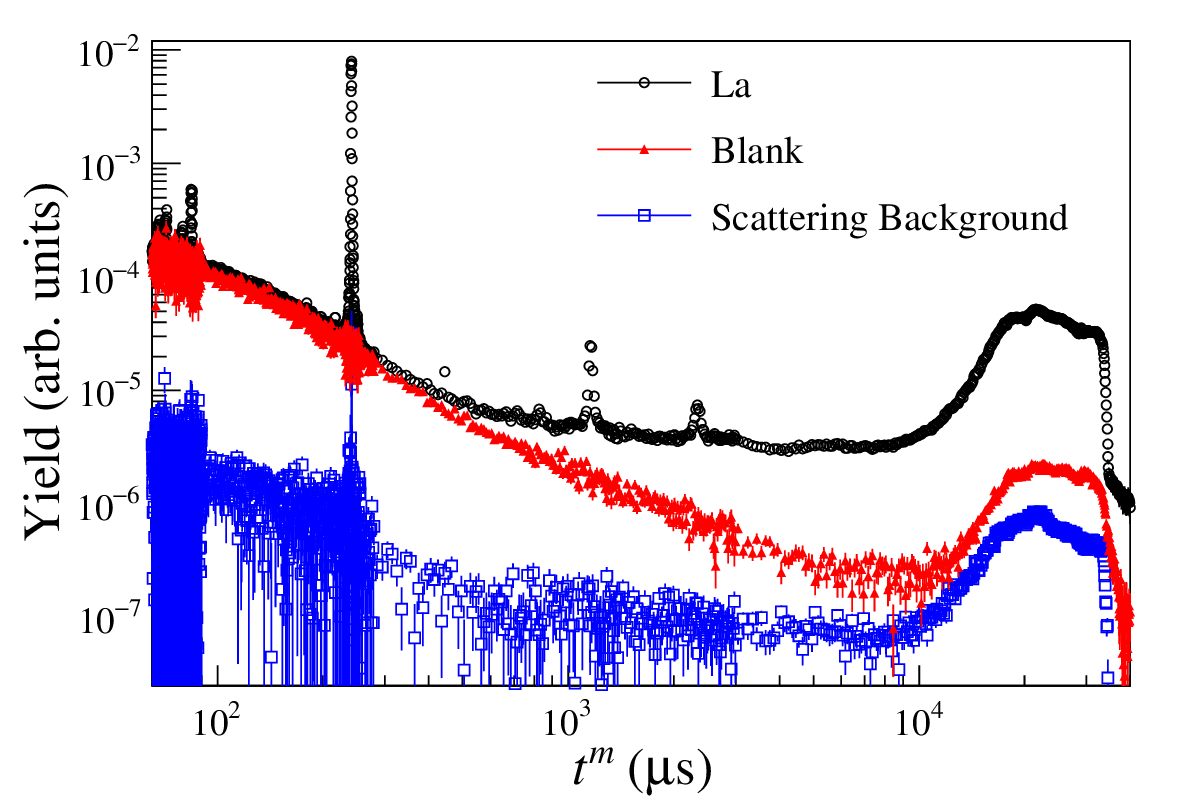}
\caption{\label{BGcor}Yields of La, blank, and scattered background after the frame-overlap subtraction.}
\end{figure}

Sample independent and sample scattered backgrou-nds were subtracted using blank and carbon measurements. The corrected yield was obtained as:
\begin{eqnarray}
\label{eq:scatcor}
Y'_\textrm{La}(t^{m})&=&(Y_\textrm{La}(t^{m})-Y_\textrm{Blank}(t^{m})) \\
&-&\frac{n_\textrm{La}\tilde{\sigma}_\textrm{s,La}(t^{m})}{n_\textrm{Carbon}\tilde{\sigma}_\textrm{s,Carbon}(t^{m})}(Y_\textrm{Carbon}(t^{m})-Y_\textrm{Blank}(t^{m})), \nonumber
\end{eqnarray}
where $Y$ is the yield after the frame-overlap correction; $n$ is the areal density; $\tilde{\sigma}_\textrm{s}$ is the scattering cross section considering the resolution function of ANNRI obtained by Kino et al.~\cite{Kino}. The second term of right-hand side of Eq. (\ref{eq:scatcor}) represents the sample-scattered background. In stead of applying the scattering cross section, the number of scattering events obtained from the PHITS simulation~\cite{PHITS} was used to consider the self-shielding effects on the scattering events. JENDL-4.0 was used as the input nuclear data. Figure~\ref{BGcor} shows yields of La, blank, and sample-scattered background.

The incident neutron flux, $\phi(t^{m})$, was obtained from the boron sample measurement by the same analysis as above. The self-shielding and multiple scattering effects were corrected by simulations using PHITS~\cite{PHITS}. 

\subsubsection{Results of capture cross section}

\begin{figure}[htbp]
\centering
\includegraphics[clip,width=8.5cm]{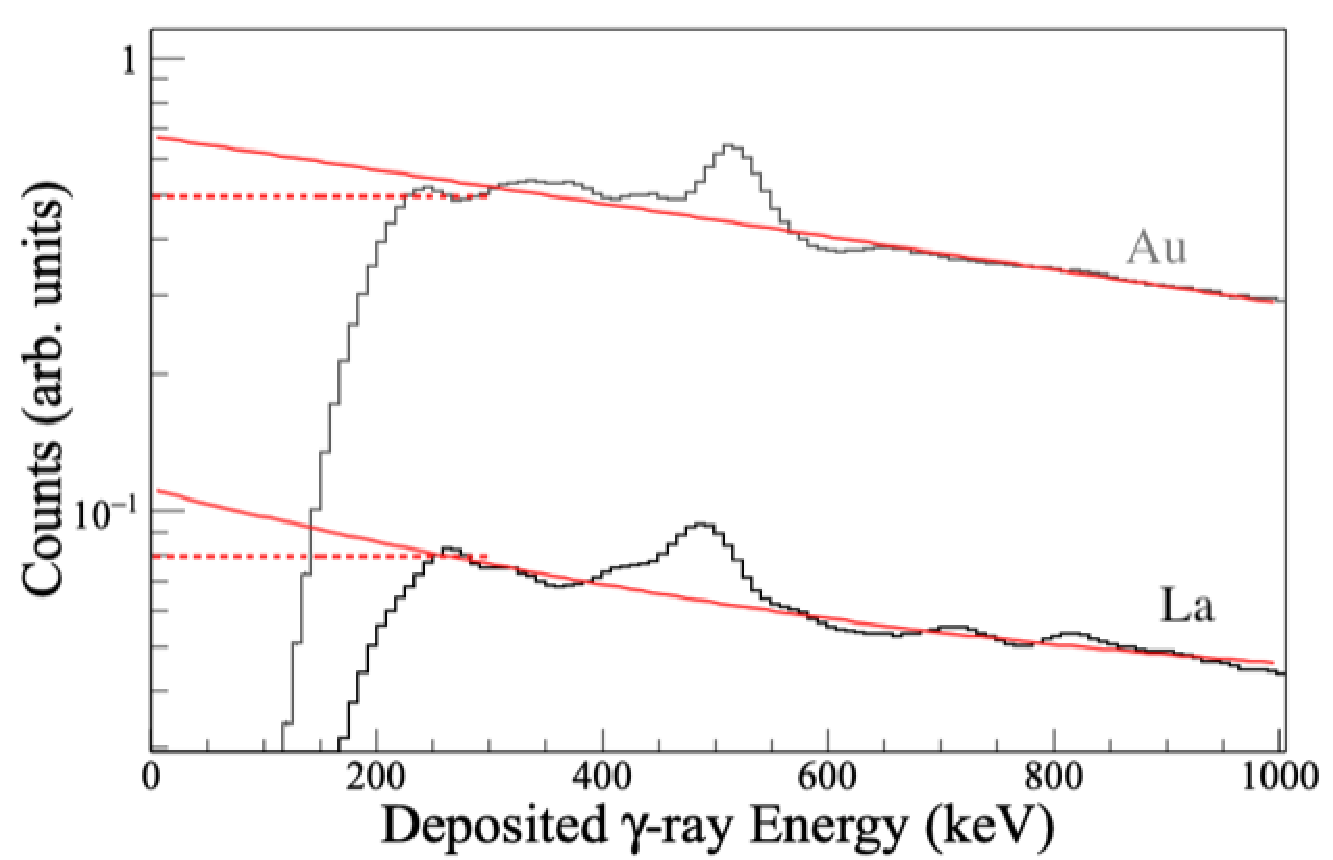}
\caption{\label{PHWTerr}Deposited $\gamma$-ray counts (histograms) for La and Au around the lower energy limit of the NaI detecter. The solid and dotted lines are the fitting results by the exponential function and constant. The uncertainty of PHWT was obtained from the difference between the integral values calculated with two extrapolated lines below the lower energy limit.}
\end{figure}

The reduced capture cross section was obtained as
\begin{equation}
    \tilde{\sigma}_\textrm{cap}(t^{m})=\frac{C_\textrm{PHWT}S_\textrm{Au}}{n_\textrm{La}S_\textrm{La}\phi(t^{m})}Y'_\textrm{La}(t^{m}),
\end{equation}
where $S_\textrm{La}$ and $S_\textrm{Au}$ are the surface area of La and Au samples; $C_\textrm{PHWT}$ is the correction factor of PHWT. The correction factor and uncertainty of PHWT are attributed to the difference in the $\gamma$-ray energy distributions of Au and La. To estimate the distribution of $\gamma$-ray counts below the detection limit of $\gamma$-ray energy, 300~keV, the method described in Ref.~\cite{Endo_Ta} was adopted. This method assumes $\gamma$-ray counts below the detection limit by fitting using exponential function or constant. Figure~\ref{PHWTerr} shows the deposited $\gamma$-ray counts for La and Au near the lower limit after subtracting those of blank, and the fitting results by the exponential function and constant. The correction factor was deduced from the ratio of analyzed events to all events for La and Au. The uncertainty was calculated from the difference between the integrals of two extrapolated lines below the detection limit. The correction factor, $C_\textrm{PHWT}$, and the uncertainty, $R_\textrm{PHWT}$, of PHWT were obtained as $C_\textrm{PHWT}=0.996$ and $R_\textrm{PHWT}=0.45\%$. 

\begin{figure}[htbp]
\centering
\includegraphics[clip,width=8.5cm]{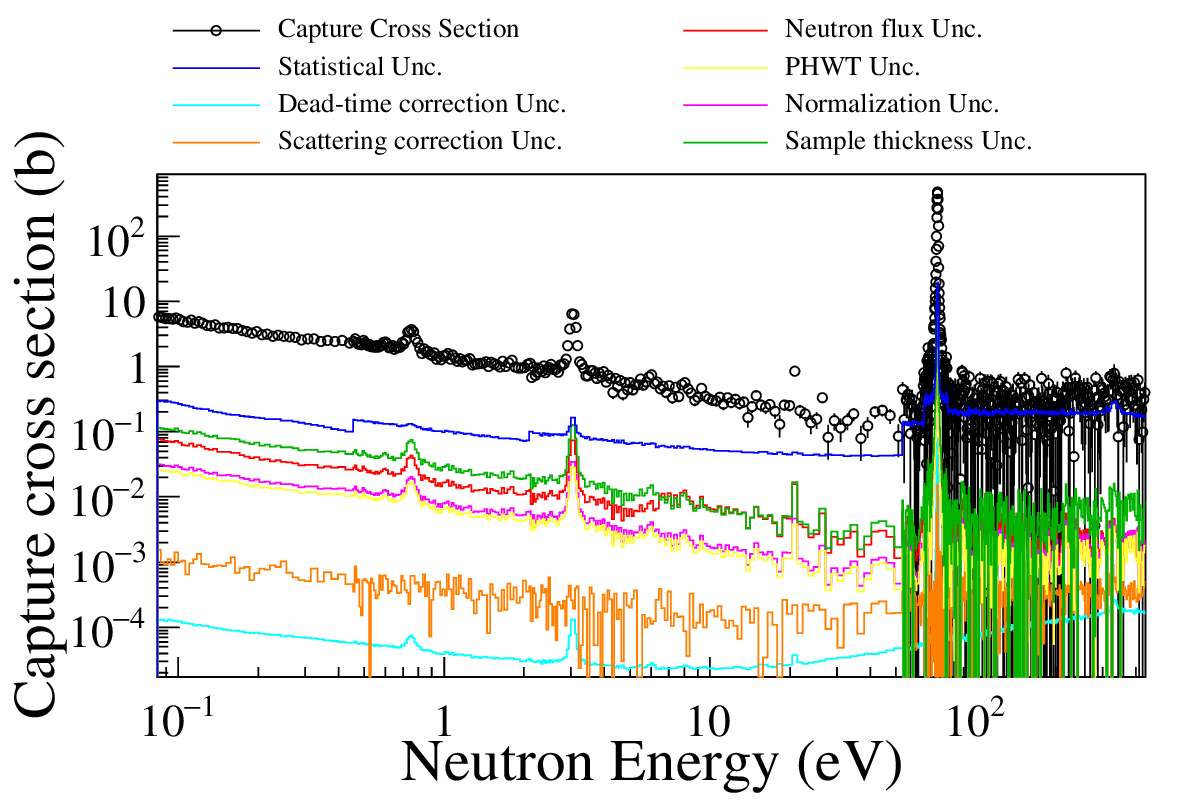}
\caption{\label{capcro}Reduced capture cross section of La with various systematic uncertainties. The impurity was not corrected.}
\end{figure}

The reduced capture cross section is shown in Fig.~\ref{capcro} with various systematic uncertainties. The normalization uncertainty was deduced from the uncertainty of the Au yield at the thermal neutron energy including statistical, dead-time, scattering, and self-shielding uncertainties.

\subsection{Resonance analysis and results}
In this study, a large deviation between the present reduced total cross section and JENDL-5 was observed around 100 eV as shown in Fig.~\ref{totcross2}. This discrepancy suggests that not only the resonance parameters but also the scattering radius should be changed. Therefore, the resonance parameters and spin-dependent scattering radius were fitted using REFIT. Fitting was made for the resonance energy, neutron width, and gamma width for the negative,
0.75-eV, and 72-eV resonances and the resonance energy and neutron width for the 1180-eV resonance. The spin, $J$, of the 72-eV resonance was used to be 3 instead of 4 as recorded in JENDL-5~\cite{JENDL} based on the suggestion from the angular correlation measurement~\cite{Okudaira}. The other resonance parameters were fixed to those in JENDL-5. The Doppler broadening was considered by the free gas model. The resolution function in ANNRI~\cite{Kino} was applied. 

The fitting regions were from 0.5 to 74 eV and from 0.5 to 1500 eV for the capture cross section and transmission, respectively. The region below 0.5~eV was not used for the fitting because the measured data were affected by gadolinium contained in the sample as an impurity. Alternatively, the thermal capture cross section of 9.25 b, which is the average value of three measurements with the activation method in the 2010s~\cite{Priyada,Farina,VanDo}, was taken into account in the fitting. The uncertainty was adjusted to have the same weight as that of the present transmission result.

\begin{figure*}[htbp]
  \begin{minipage}{0.5\hsize}
  \begin{center}
  \subfloat[Overall region.]{\includegraphics[width=85mm]{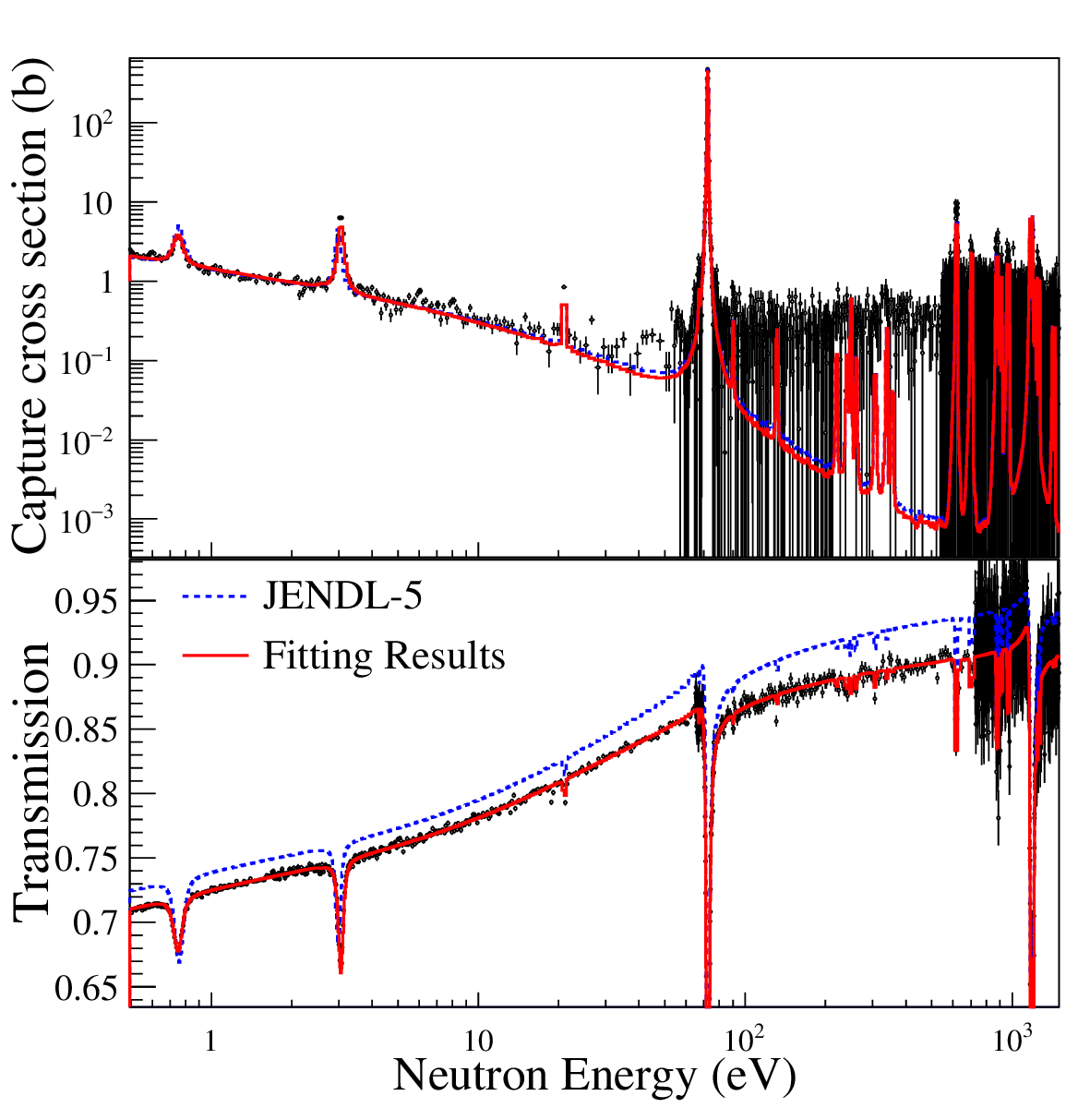}}
  \end{center}
 \end{minipage}
 \begin{minipage}{0.5\hsize}
  \begin{center}
    \subfloat[Around the 0.75-eV resonance.]{\includegraphics[width=85mm]{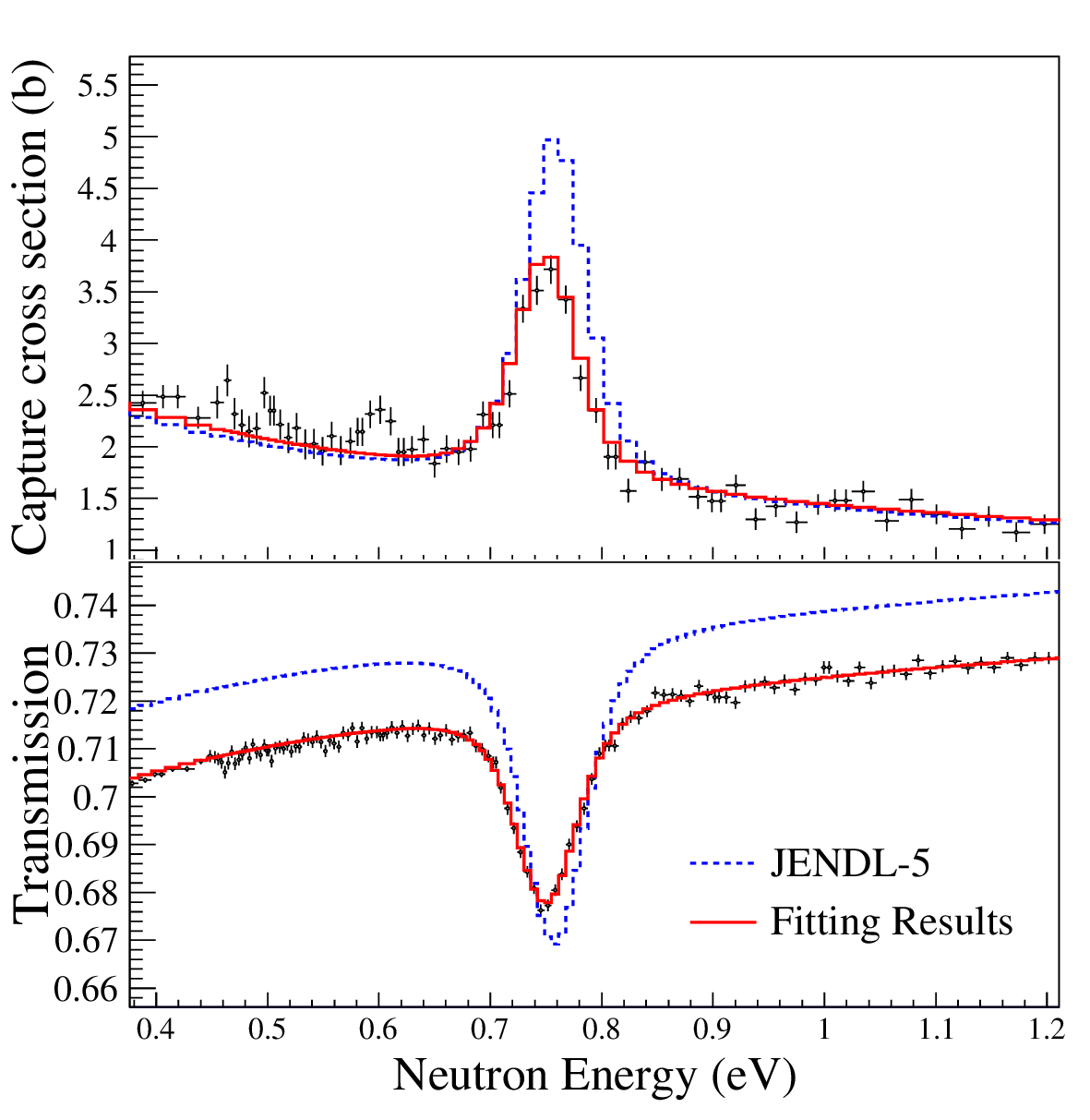}}
  \end{center}
 \end{minipage}
 \\
 
 \begin{minipage}{0.5\hsize}
 \begin{center}
   \subfloat[Around the 72-eV resonance.]{\includegraphics[width=85mm]{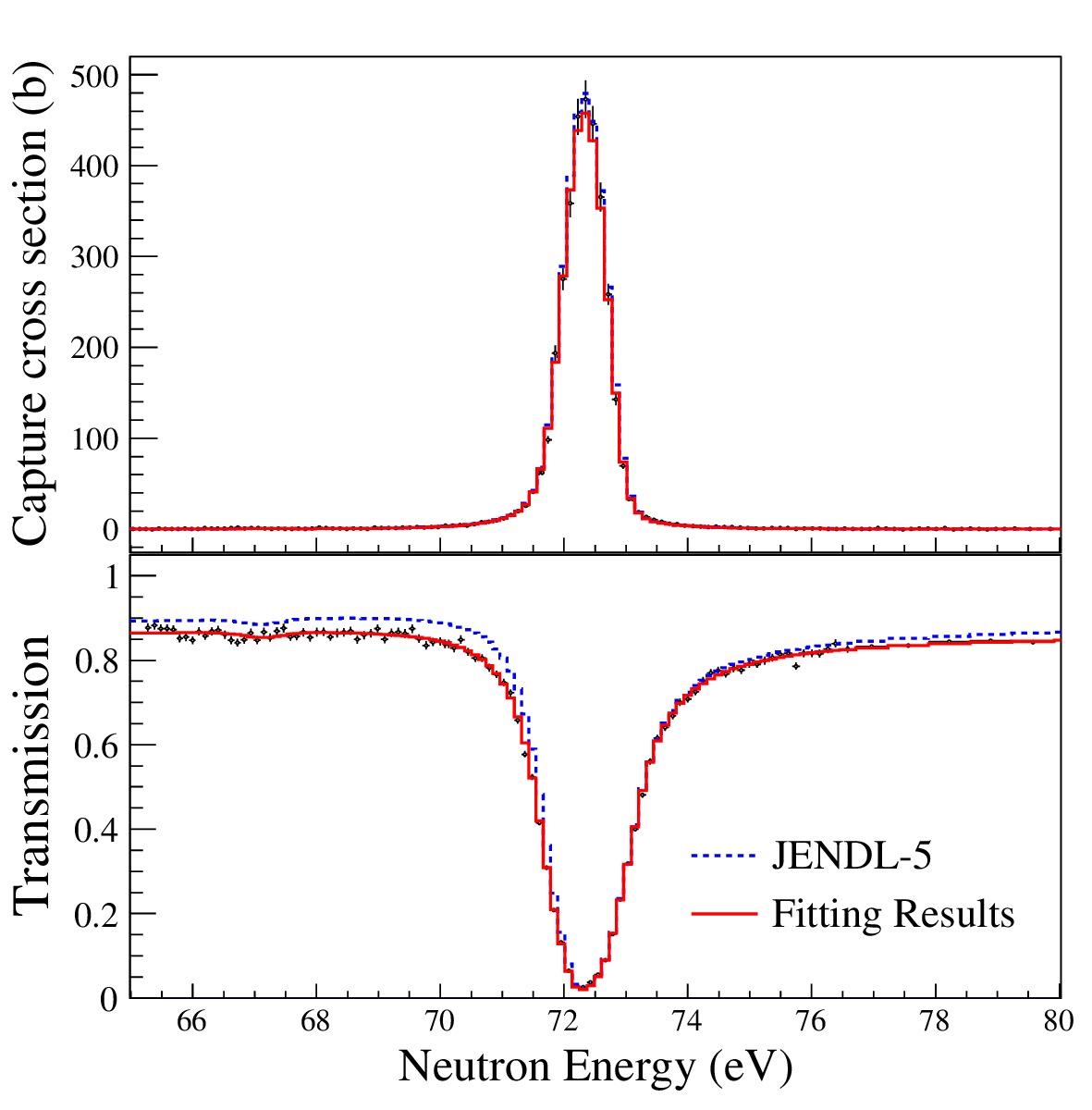}}
 \end{center}
 \end{minipage}
 \begin{minipage}{0.5\hsize}
 \begin{center}
    \subfloat[Around the 1179-eV resonance.]{\includegraphics[width=80mm]{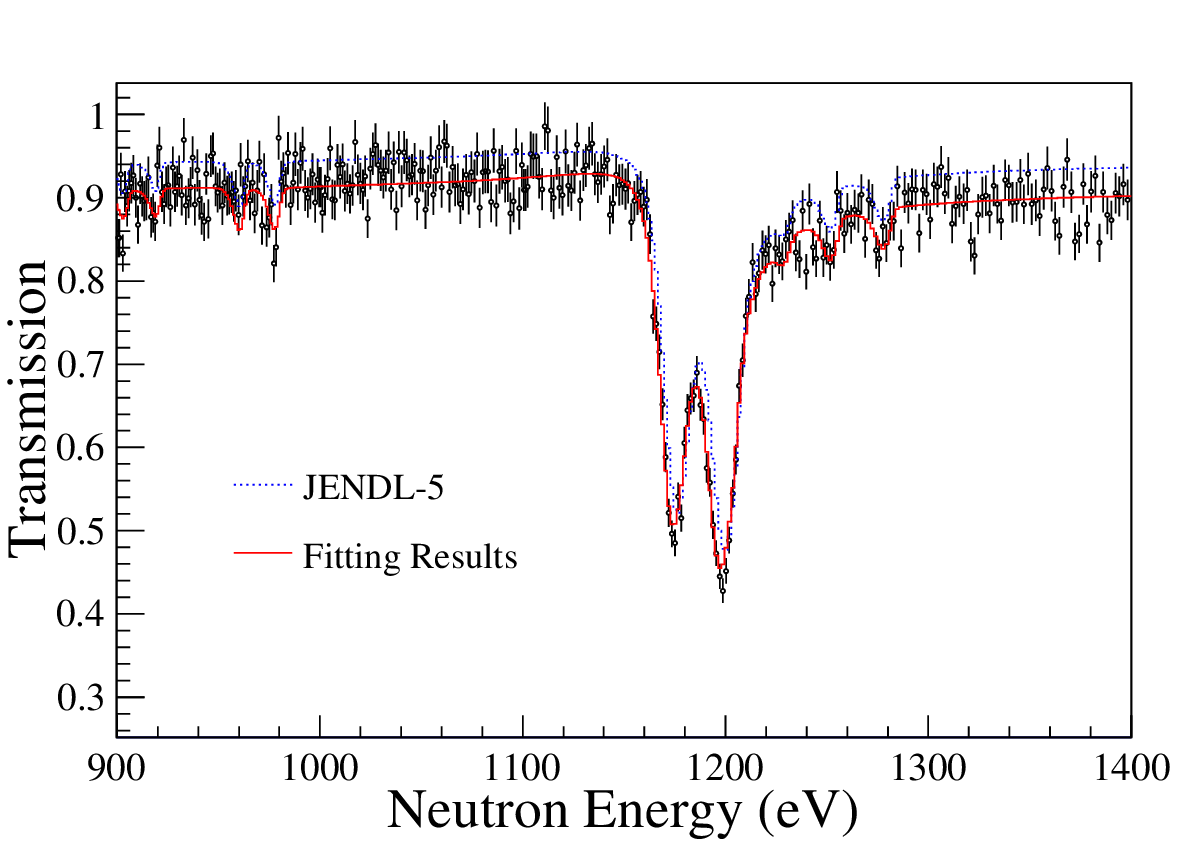}}
 \end{center}
 \end{minipage}
 \caption{\label{Fit} Reduced capture cross section (top) and transmission (bottom) with the fitting results and JENDL-5. Only transmission was fitted around 1179-eV resonance. The capture cross section and transmission below 0.5~eV and the capture cross section around 0.6~eV were not considered in the fitting because of the impurities.}
\end{figure*}

Figure \ref{Fit} displays the fitting results and JENDL-5. The peaks around 0.6~eV seems to be derived from the erbium-167 of impurity, but the shape of the resonance did not match the experimental results. Therefore, this region was not considered in the fitting. The spin-dependent scattering radii were obtained to be $R'_{J=4}$ $=$ $6.13\pm0.10$~fm and $R'_{J=3} = 5.54\pm0.18$~fm, and the averaged scattering radius resulted in $R' = 5.89\pm0.09$~fm which was calculated as
\begin{equation}
    R'=\sqrt{g_{J=4}R_{J=4}^{'2}+g_{J=3}R_{J=3}^{'2}},
\end{equation}
where $g$ is the spin factor, $g=(2J+1)/(2(2I+1))$; $I$ is the nuclear spin. The obtained resonance parameters are listed in Table~\ref{respara}. The past data are taken from Terlizzi et al.~\cite{nTOF}, Alfimenkov et al.~\cite{Alfimenkov}, Shwe et al.~\cite{Shwe}, and Harvey et al.~\cite{Harvey}. 

\begin{table*}[htbp]
\centering
\caption{\label{respara} Resonance parameters of $^{139}$La. The spin, $J$, and the neutron orbital angular momentum, $l$, of resonances are taken from JENDL-5 except for the 72-eV resonance. $\Gamma_\gamma$ and $\Gamma_\textrm{n}$ are the gamma and neutron widths, respectively.}
 \begin{tabular}{cc|ccc}
 $J/l$ & Reference & Resonance Energy [eV] & $\Gamma_\gamma$ [meV] & $g\Gamma_\textrm{n}$ [meV] \\ \hline
 4/0  & Present work & $-38.8\pm0.4$  & $60.3\pm0.5$ &  $346\pm10$ \\
  & JENDL-5 & -48.63 & 60.1 & 584.4 \\
  & ENDF/B-VIII.0  & -48.63 & 62.2 & 572.1  \\ \hline

   4/1      & Present work& $0.750\pm0.001$ & $41.6\pm0.9$ & $(3.67\pm0.05)\times10^{-5}$ \\ 
  &Terlizzi et al.       & $0.758\pm0.001$     & $40.1\pm1.9$     & $(5.6\pm0.5)\times10^{-5}$    \\
  & Alfimenkov et al. & $0.75\pm0.01$ & $45\pm5$ & $(3.6\pm0.3)\times 10^{-5}$ \\
  &Shwe et al.       & $0.734\pm0.005$      & $40\pm5$     & $(3.67\pm0.22)\times10^{-5}$    \\
  &Harvey et al.       & $0.752\pm0.011$      & $55\pm10$     & $(4\pm1)\times10^{-5}$    \\
 &    JENDL-5 & 0.758 & 40.1 & $5.60\times10^{-5}$ \\
& ENDF/B-VIII.0  & 0.734 & 45.0 & $3.65\times 10^{-5}$  \\ \hline

  3/0    & Present work & $72.30\pm0.01$ & $64.1\pm3.0$ & $ 13.1\pm0.7$ \\
 &Terlizzi et al.  & $72.30\pm0.05$      & $75.6\pm2.2$     & $11.8\pm0.5$    \\
   &Shwe et al.       & $72.3\pm0.1$      & $56.5\pm1.7$     & $13.8\pm0.2$    \\
 &    JENDL-5 & 72.3 & 75.6 & 11.76 \\
& ENDF/B-VIII.0  & 72.3 & 52.6 & 15.15 \\  \hline

  3/0    & Present work& $1179.0\pm0.2$ & $91.6$ (fixed) & $941\pm35$ \\
  & Terlizzi et al.  & $1181\pm4$ & $91.6\pm2.7$ & 923 (fixed) \\
  & Shwe et al.  & $1182\pm2$ & 56.5 (fixed) & $920\pm10$ \\
   &    JENDL-5 & 1181 & 91.6 & 923 \\
& ENDF/B-VIII.0  & 1178.7 & 50.6 & 932.5 \\

\end{tabular}
\end{table*}

\section{Discussion}
\subsection{Cross section and positive resonance parameters}
\begin{figure}[htbp]
\centering
\includegraphics[clip,width=8.2cm]{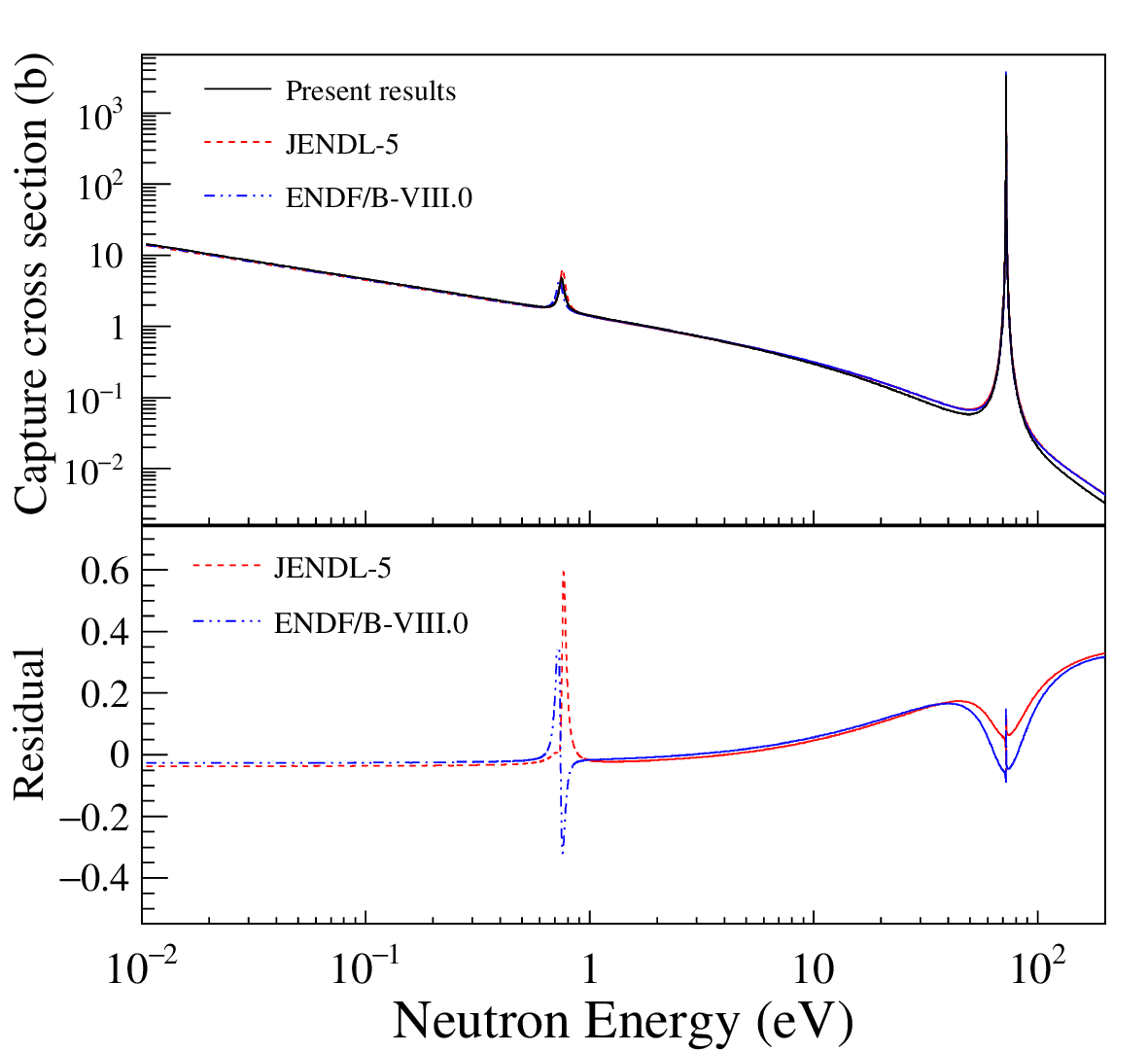}
\caption{\label{capcro_eval} Capture cross section calculated from the resonance parameters of the present results and the evaluated libraries (top) and residual (bottom).}
\end{figure}

\begin{figure}[htbp]
\centering
\includegraphics[clip,width=8.5cm]{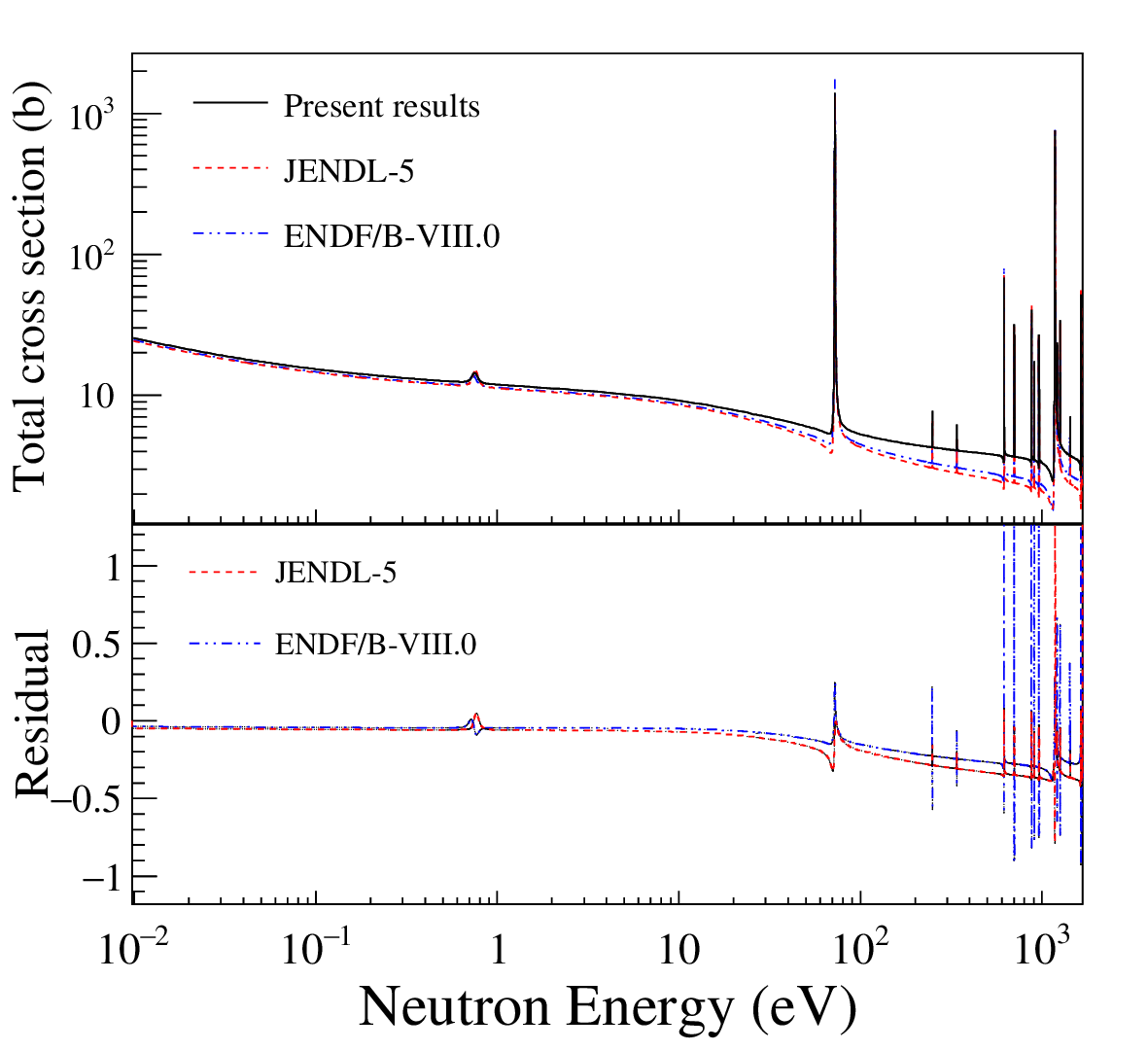}
\caption{\label{totcro_eval} Total cross section calculated from the resonance parameters of the present results and the evaluated libraries (top) and residual (bottom).}
\end{figure}

Figures \ref{capcro_eval} and \ref{totcro_eval} demonstrate the cross section calculated from the resonance parameters of the present results, $\sigma_\textrm{present}$, and the evaluated libraries, $\sigma_\textrm{evaluation}$, with the residual defined as
\begin{equation}
\label{residual}
\frac{\sigma_\textrm{evalation}(E_\textrm{n})-\sigma_\textrm{present}(E_\textrm{n})}{\sigma_\textrm{present}(E_\textrm{n})}.
\end{equation}
Regarding the 0.75-eV resonance, the peak of the present total and capture cross sections was almost same as those of ENDF/B-VIII.0, but it was 60\% smaller than that of JENDL-5. The total cross section around 100~eV was about 20\% different from the evaluated data due to the change of the scattering radius. The thermal neutron capture cross section was calculated to be $9.28$ b from the obtained resonance parameters. This value was a little larger than those in the evaluated libraries, 8.94~b and 9.04~b in JENDL-5 and ENDF-B/V.III.0, which were considered too small based on the results of the recent activation measurements listed in Table~\ref{thermalcross}.


\begin{table}[htbp]
\centering
\caption{\label{thermalcross} Thermal neutron capture cross section.}
 \begin{tabular}{c|c}
 Reference & Cross Section [b] \\ \hline
 Present work & $9.28$\\
 Priyada et al. (2017) ~\cite{Priyada} & $9.24\pm0.25$ \\
Priyada et al. (2017) ~\cite{Priyada}& $9.28\pm0.37$ \\
 Nguyen et al. (2014) ~\cite{VanDo}& $9.16\pm0.36$ \\
 Farina et al. (2013) ~\cite{Farina}& $9.16\sim9.30$ \\
\end{tabular}
\end{table}

The neutron width of the 0.75-eV resonance was different from that of Terlizzi et al.~\cite{nTOF}, but in good agreement with that of Alfimenkov et al.~\cite{Alfimenkov}, Shwe et al.~\cite{Shwe} and Harvey et al.~\cite{Harvey}. Terlizzi et al. evaluated the parameter from the neutron capture cross section, while the other studies obtained it from the neutron transmission. The present results were evaluated from both the transmission and capture cross section experiments, both of which are in good agreement with the fitting results as shown in Fig.~\ref{Fit}(b).

To estimate influence on the measurement of the power distribution in nuclear reactor, the averaged value of the cross section was calculated from the cross section of the present results and JENDL-5. The value is defined as:
\begin{equation}
\overline{\sigma}=\int_0^{10 keV}\sigma(E)\psi(E)dE/\int_0^{10 keV}\psi(E)dE,
\end{equation}
where $\psi(E)$ is the neutron spectrum in a pressurized water reactor taken from Ref.~\cite{ReactorSpec}. The averaged values were obtained to be $2.15$~b and $2.13$~b from the present results and JENDL-5, respectively. It was found that the change of resonance parameters only affects the derivation of power distribution by about 1\%.

\subsection{Negative resonance parameter}

Negative resonance parameters were different from the evaluated libraries. Figure~\ref{capcro_eval} indicates that the change of the negative resonance parameter has little effect on the capture cross section in the energy range below 1 eV. Change of about 20\% was observed in the energy range above 10~eV, but it is difficult to measure the capture cross section in this range due to the low cross section. On the other hand, Fig.~\ref{totcro_eval} presents the parameters of the negative resonance have a significant effect on the energy dependence of the total cross section. Therefore, it is necessary to fit the transmission over a wide energy range to derive the negative resonance parameters. The difference from the evaluation might be caused by the insufficient consideration of this energy dependence of transmission.

Past evaluated values of negative resonance are listed in Table \ref{negres}. The parameters in Mughabghab (1981)~\cite{Mughabghab1981} were changed significantly from those in Mughabghab (1973)~\cite{Mughabghab1973}. The older one is relatively close to the present results. It is noted that ENDF/B-VII.0 cited Mughabghab (2006)~\cite{Mughabghab2006} while ENDF/B-VI.0 did Mug-habghab (1973). The measurements added up to the update from Mughabghab (1973) were several thermal neutron capture cross sections by the activation method, two total cross sections, and one capture cross section. The thermal neutron capture cross section was little changed from $9.0\pm0.3$~b to $8.93\pm0.04$~b. Since the capture cross section measurement~\cite{Musgrove} was performed in the energy range above 2.7 keV, it is not seems to affect the evaluation of the negative resonance parameters. Two total cross section measurements are conducted in the energy ranges above 72.3~eV~\cite{Hacken} and 1~keV~\cite{Camarda}, respectively. The reason for such change of resonance parameters in Mughabghab (1981) was not known, but it might be difficult to accurately obtain the negative resonance parameters just from the addition of these experimental data.

\begin{table*}[htbp]
\centering
\caption{\label{negres} Resonance parameters of the negative resonance in each evaluation.}
 \begin{tabular}{c|ccc}
 Reference & Resonance Energy [eV] & $\Gamma_\gamma$ [meV] & $g\Gamma_\textrm{n}$ [meV] \\ \hline
  Present work & $-38.8\pm0.4$  & $60.3\pm0.5$ &  $346\pm10$ \\
 JENDL-5 (2023) & -48.63 & 60.1 & 584.4 \\
 Mughabghab (2019) & -48.63 & (60.94) & 592.7  \\ 
 ENDF/B-VIII.0 (2018) & -48.63 & 62.2 & 572.1  \\ 
 ENDF/B-VII.0 (2006) & -48.63 & 62.2 & 572.1  \\ 
Mughabghab (2006) & -48.63 & (62.2) & 571.8  \\ 
ENDF/B-VI.0 (1990) & -37.5 & 50.1 & 352.1  \\ 
Mughabghab (1981) & -48.63 & (62.2) & 585.8  \\ 
Mughabghab (1973) & -37.5 & (56.5) & 352 \\ 

\end{tabular}
\end{table*}

\begin{figure}[htbp]
\centering
\includegraphics[clip,width=8.5cm]{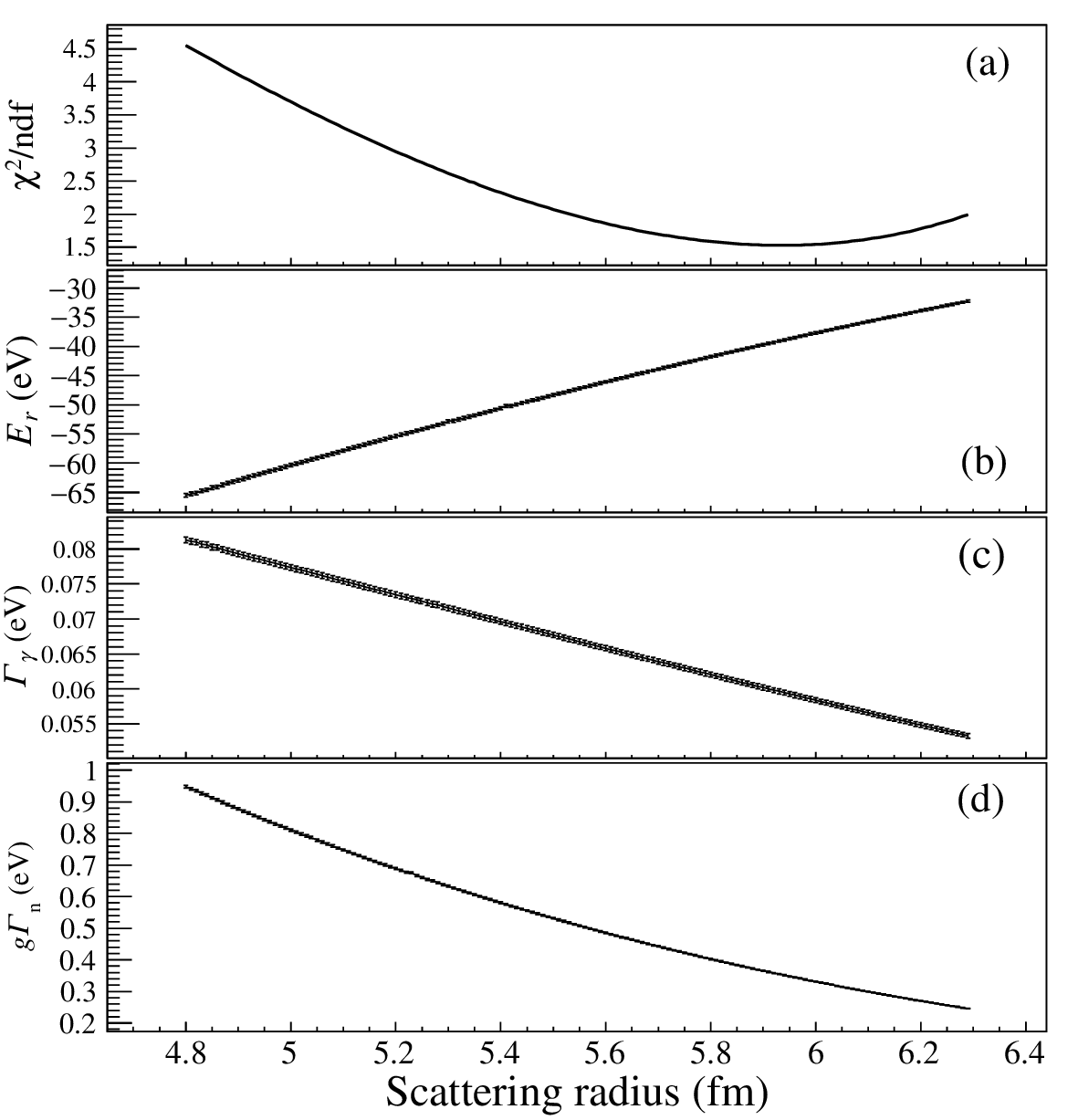}
\caption{\label{Para_Rdep} Reduced chisquare $\chi^2/$ndf (a) and negative resonance parameters, resonance energy (b), gamma width (c), and neutron width (d), for each scattering radius. The neutron width is multiplied by the statistical factor, $g$.}
\end{figure}

\begin{figure}[htbp]
\centering
\includegraphics[clip,width=8.5cm]{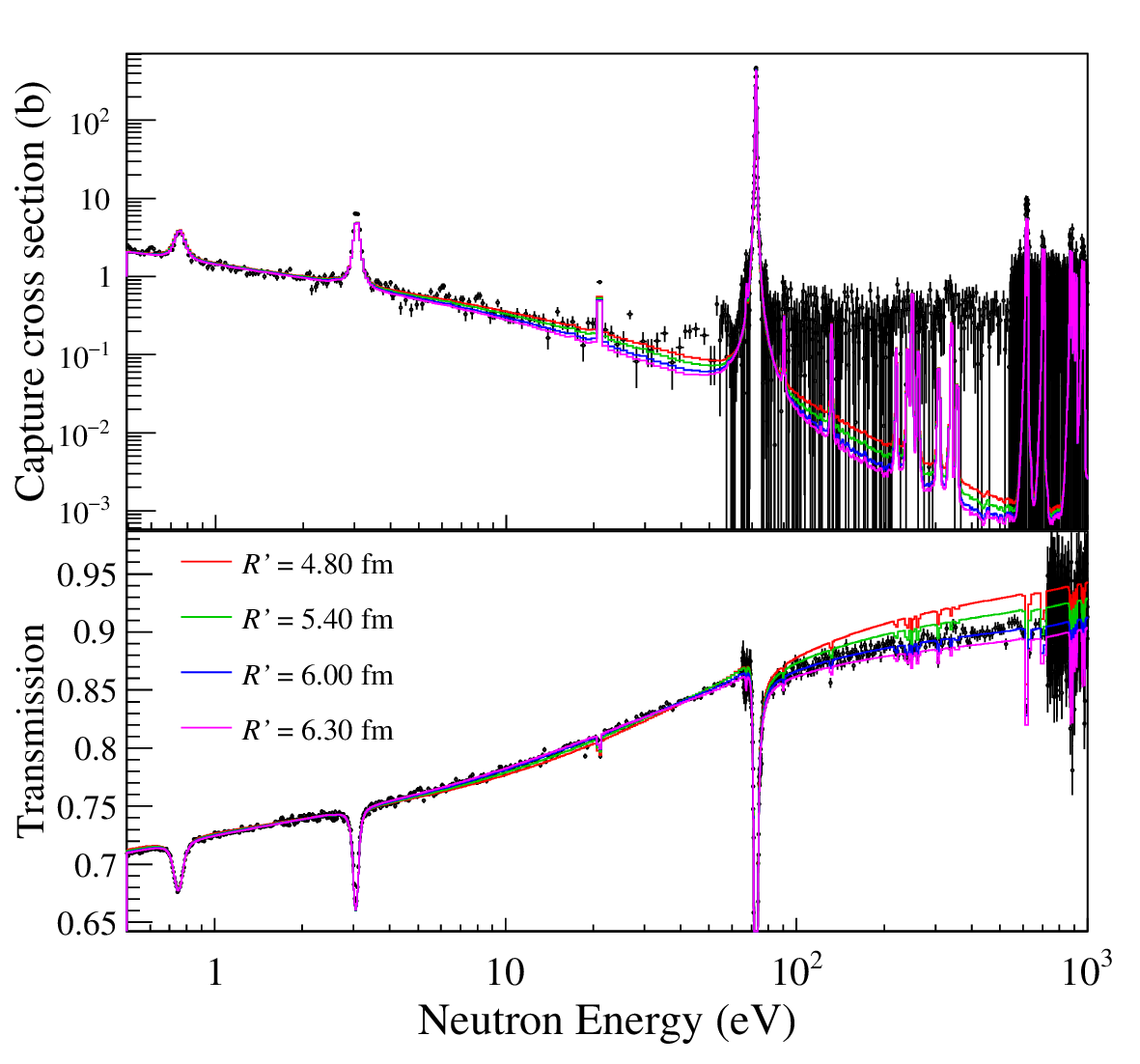}
\caption{\label{Fit_Rdep} Fitting results of capture cross section (top) and transmission (bottom) for each scattering radius.}
\end{figure}

The relationship between the negative resonance parameters and the scattering radius $R'$ is discussed. The scattering radius was changed from 4.80 fm to 6.30 fm in 0.01 fm steps, and the resonance parameters including negative resonance were fitted. Figure~\ref{Para_Rdep} shows the resonance parameters of the negative resonance for each scattering radius along with the value of the reduced chisquare, $\chi^2/$ndf. The obtained parameters were a little different from those in Table~\ref{respara} because the spin-dependent scattering radius was not considered in this analysis. Figure~\ref{Fit_Rdep} displays the fitting results for each scattering radius and indicates that the fitting was not good for $R'$ value being far from 6~fm, such as $R'=4.80$~fm and $R'=6.30$~fm, so that the value of the reduced chisquare became large.  The reduced chisquare in Fig.~\ref{Para_Rdep} pointed out that the fitting results could not reproduce the experimental results with only an adjustment of the negative resonance parameters when the $R'$ value is far from 6~fm. In other words, to reproduce the experimental results, both the $R'$ value and the negative resonance parameters needed to be adjusted.

\begin{figure}[htbp]
\centering
\includegraphics[clip,width=8.5cm]{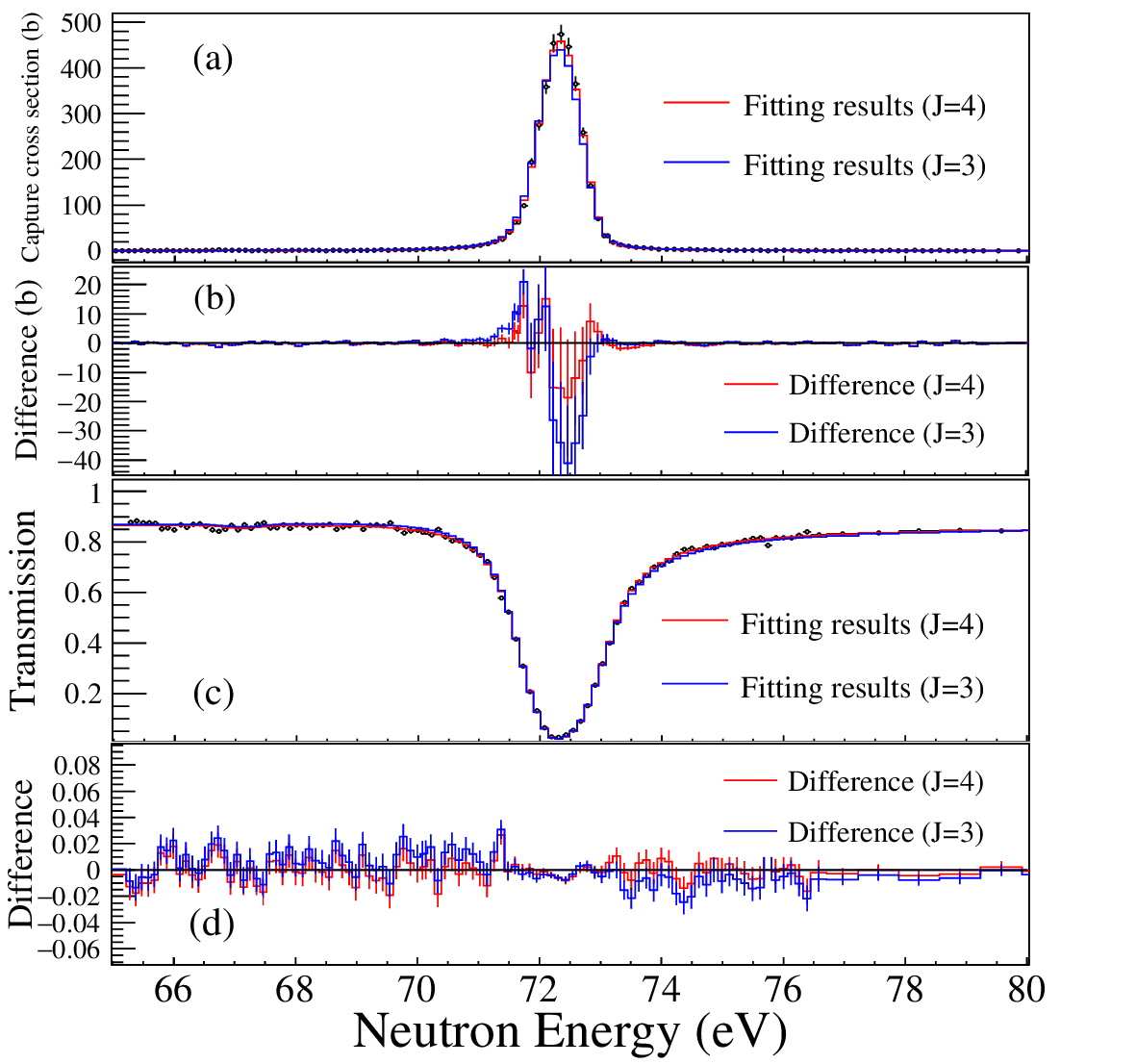}
\caption{\label{NegJ3} Comparison of fitting results in the case that the negative resonance spin is assumed to be $3$ or $4$. (a): capture cross section; (b): difference of capture cross section between the fitting and experimental results; (c): transmission; (d): difference of transmission. }
\end{figure}

The value of spin $J$ for the negative resonance which affects most strongly on the positive energy region was assumed as 4 in aforementioned analysis. Here we discuss the case that the spin of the negative resonance is 3. The resonance parameters were evaluated using REFIT in the same method with the negative resonance spin as 3. Figure~\ref{NegJ3} compares the fitting results in the cases of $J=3$ and $J=4$ around the 72-eV resonance. Difference was calculated as the fitting results minus experimental results. The spin value of negative resonance has the largest influence on the shape of transmission near the 72-eV resonance due to the interference. Figure~\ref{NegJ3} indicates that in the case of $J=3$ the fitting result of the transmission is larger than the experimental result at the low energy-side, while the one is smaller than that at high energy-side. Therefore, in the case of $J=3$, the cross section calculated from resonance parameters could not reproduce the measurement results well. Furthermore, the average level spacing was evaluated to be $\braket{D}=252$~eV~\cite{nTOF}. The first resonance of $J=3$ is located at 72-eV, while the one of $J=4$ is done at $994$~eV. Therefore, the first negative resonance is more likely the $J=4$ resonance, considering the resonance spacing. 


The parity-violation amplitude is predicted by the s-p mixing model~\cite{Flambaum} to be
\begin{equation}
p=2\sum_{s}\frac{V_\textrm{sp}^{J}}{E_\textrm{s}-E_\textrm{p}}\sqrt{\frac{\Gamma_\textrm{n,s}}{\Gamma_\textrm{n,p}}},
\end{equation}
where $V_\textrm{sp}^{J}$ is the individual weak matrix element. In $^{139}$La case, the value of $p=(9.56\pm0.35)\times 10^{-2}$ was established~\cite{Michel}. The weak matrix element was calculated to be $-2.16$ and $-1.70$ meV using resonance parameters in JENDL-5 and present results, respectively. The weak matrix element was changed by 25\%, which was much larger than the accuracy of the parity-violation measurement, $4\%$. The accuracy of the resonance parameters used to interpret the results of the angular correlation measurements based on the s-p mixing model would be important in the future.

\subsection{Potential scattering radius of $^{139}$La}
\begin{figure}[htbp]
\centering
\includegraphics[clip,width=8.5cm]{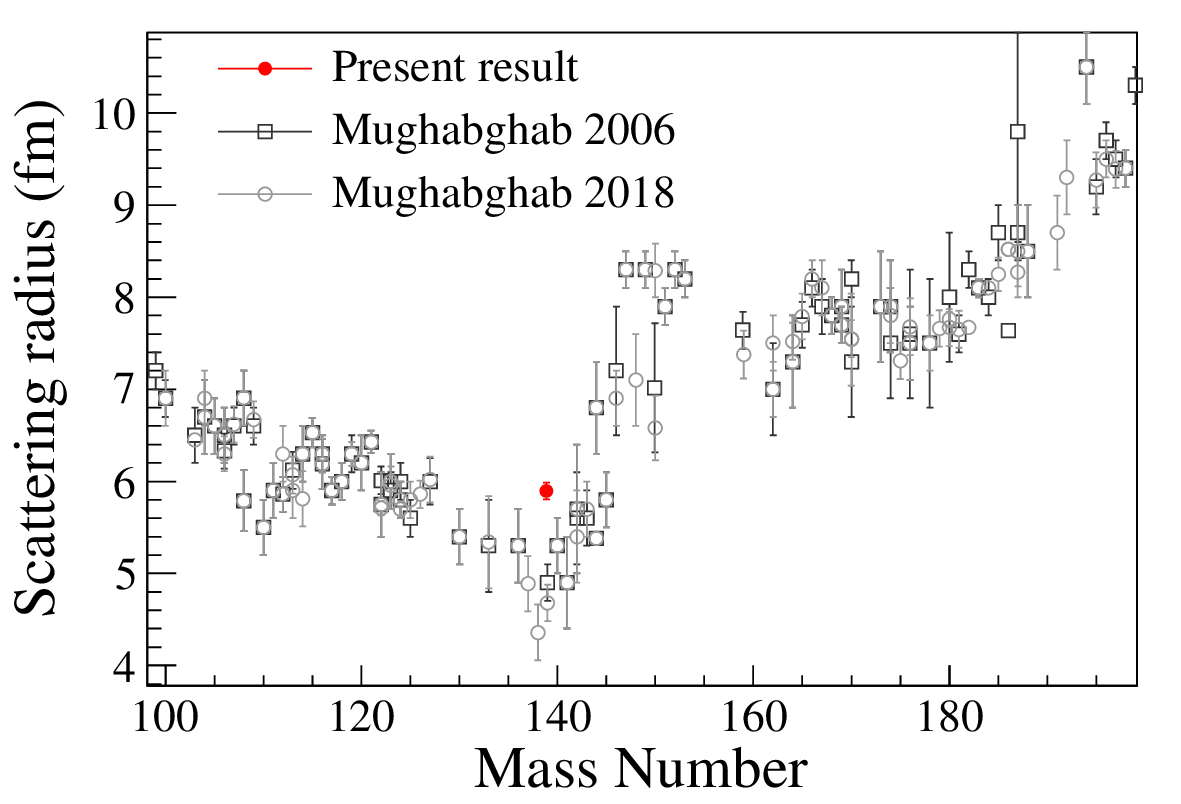}
\caption{\label{R_A_Dep} Mass number dependence of scattering radius taken from Mughabghab ~\cite{Mughabghab2006,Mughabghab}.} 
\end{figure}

The present scattering radius, 5.94~fm on average, is larger than 4.8~fm recorded in JENDL-5.  Shwe et al.~\cite{Shwe} estimated the scattering radius as $R'=5.2\pm1.0$~fm by a curve fit of the transmission around the 72-eV resonance. In their analysis, the spin dependence of the nuclear radius was not considered. Since the 72-eV resonance is $J=3$, it is appropriate to compare the $R'$ from Shwe et al. with the $R'_{J=3}=5.54\pm0.18$~fm in the present analysis, and they are in good agreement. Figure \ref{R_A_Dep} shows the mass number dependence of scattering radius taken from Mughabghab~\cite{Mughabghab2006,Mughabghab} with the present results. The scattering radii of nuclei near $^{139}$La indicates that the value of 5.89~fm does not deviate significantly from those of near nuclei although it seems a little large. 

Knopf and Waschkowski~\cite{Knopf} determined the scattering radius as $R'=4.6\pm0.2$ fm by a Christiansen-Filter Technique~\cite{Koester,Koes}. The scattering radius in Mugh-abghab and JENDL-5 was evaluated based on this results. In the Christiansen-Filter Technique, the coherent scattering length at the slow neutron energy can be obtained, and the scattering radius is deduced by subtracting the resonance effect from the scattering length. The scattering radius was calculated as~\cite{Mughabghab}
\begin{equation}
\label{eq:R}
R'=\frac{A}{A+1}b+2.277\times 10^3\left(\frac{A+1}{A}\right)\sum_{j}\frac{\Gamma_{\textrm{n},j}^0}{E_{0,j}},
\end{equation}
where $A$ is the mass number; $b$ is the bound coherent scattering length; $E_{0,j}$ is the resonance energy; $\Gamma_{\textrm{n},j}^0$ is the reduce neutron width defined as 
\begin{equation}
\Gamma_{\textrm{n},j}^0=\sqrt{\frac{1}{|E_{0,j}|}}\Gamma_{\textrm{n},j}.
\end{equation}
The second term in Eq. (\ref{eq:R}) represents the resonance effect. The negative resonance effect was calculated using the present results and the Mughabghab parameters~\cite{Mughabghab1981}, which Knopf and Waschkowski employed, were $-5.8$ and $-7.1$~fm, respectively. Therefore, if the present result for negative resonance is used, the scattering radius obtained from the coherent scattering length becomes $4.6+(7.1-5.8)=5.9$~fm. This result is in good agreement with the present results. Therefore, the value, 4.8 fm, recorded in JENDL-5 may be small.

\section{Conclusion}
The cross sections and resonance parameters of $^{139}$La are significant for the applications in nuclear technology and the studies in fundamental nuclear physics. In particular, the resonance parameters are required to interpret the measurement results of the angular correlation terms for verification of the s-p mixing model. The neutron total and capture cross sections were measured in J-PARC$\cdot$MLF$\cdot$ANNRI. The total cross section is in good agreement with that of past measurements, but the discrepancy with the evaluated libraries in non-resonance region around 100~eV was found. Since the scattering radius could have been estimated to be small, not only the resonance parameters but also the scattering radius was evaluated using the obtained transmission and capture cross section by the resonance analysis code, REFIT. The resonance parameters of four resonances including one negative resonance were obtained. The neutron width of the 0.75-eV resonance was different from that in the most recent measurement but consistent with that in other previous measurements. The negative resonance parameters were also different from the evaluated nuclear data libraries. The spin-dependent scattering radii were obtained to be $R'_{J=4}=6.13\pm0.10$~fm and $R'_{J=3}=5.54\pm0.18$~fm, and the averaged scattering radius was calculated as $R'=5.89\pm0.09$~fm. This result was compared with the past measurements, and it was pointed out that the scattering radius recorded in the evaluated libraries might have been estimated too small.

\begin{acknowledgements}

The authors would like to thank the staff of ANNRI for the maintenance of the germanium detectors, and MLF and J-PARC for operating the accelerators and the neutron production target. The neutron experiments at the MLF of J-PARC were performed under the user program (Proposals No. 2023P0300). This work was supported by the Neutron Science Division of KEK as an S-type research project with program number 2018S12. This work was partially supported by JSPS KAKENHI Grant No. JP20K14495 and JP21K04950.
\end{acknowledgements}

\bibliographystyle{spphys}       
\bibliography{sn-bibliography}   

\end{document}